\documentclass[aps,showpacs,showkeys,preprint]{revtex4}
\usepackage[dvips]{graphicx}
\usepackage{latexsym}
\begin{document}

\begin{flushright}
{\tt SOGANG-MP 02/07}
\end{flushright}

\begin{flushright}
\today
\end{flushright}

\title{Entropy of (2+1)-dimensional de Sitter black hole to all orders \\
in the Planck length}
 \author{Yong-Wan \surname{Kim}}
   \email{ywkim65@gmail.com}
 \affiliation{Institute of Mathematical Science and
School of Computer Aided Science, \\Inje University, Gimhae 621-749,
Korea}
 \author{Young-Jai Park}
 \email{yjpark@sogang.ac.kr}
 \affiliation{Department of Physics and Mathematical Physics Group, Sogang University, Seoul 121-742, Korea}

\begin{abstract}
We calculate the statistical entropy of a scalar field on
the background of (2+1)-dimensional de Sitter space without an
artificial cutoff considering corrections to all orders in the Planck length from a
generalized uncertainty principle (GUP) on the quantum state
density. The desired entropy proportional to the horizon
perimeter is obtained.
\end{abstract}


\maketitle

About three decades ago, Bekenstein had suggested that the entropy
of a black hole is proportional to the area of horizon from a view
point of information theory \cite{bek}. Subsequently, Hawking
showed that black hole entropy satisfies exactly the area law by
means of Hawking radiation based on quantum field theory
\cite{haw}. After their works, 't Hooft investigated the
statistical properties of a scalar field outside the horizon of
Schwarzschild black hole by introducing an artificial brick wall
cutoff \cite{tho} in order to remove the ultraviolet divergence
near the horizon \cite{gm,kkps,lz}. Recently, several
authors~\cite{li,liu,kkp} calculated the entropy of black holes to
leading order in the Planck length by using a generalized
uncertainty principle (GUP) \cite{gup,gup1} solving the
ultraviolet divergences of the just vicinity near horizon
replacing a brick wall cutoff with a minimal length. In
particular, by using a GUP we have studied the entropy of a scalar
field on the (2+1)-dimensional de Sitter (DS) black hole
background to leading order in the Planck length~\cite{kpo}
improving the previous DS works having a brick wall \cite{wlo}.
However, Yoon et. al. have very recently pointed out that since a
minimal length is actually related to a brick wall cutoff, the
entropy integral about radial component $r$ in the range near
horizon should be carefully treated for a convergent entropy
\cite{yoon}.

On the other hand, it is also well-known that the deformed
Heisenberg algebra \cite{gup} leads to a GUP showing the existence
of a minimal length \cite{kn0,kn,kpn,Moffat,spallucci,nouicer},
which originates due to the quantum fluctuation of the gravitational
field \cite{ven,mag}.
Indeed, it has been shown that the Feynman propagator displays an
exponential ultraviolet cutoff of the form of $\exp \left(
-\lambda p^2\right) $, where the parameter $\sqrt{\lambda}$
actually plays a role of a minimal length as shown later.
Moreover, quantum gravity phenomenology has been tackled with
effective models based on GUPs and/or modified dispersion
relations \cite{Cam,Sabine} containing a minimal length as a natural
ultraviolet cutoff \cite{fsp}. The essence of ultraviolet
finiteness of the Feynman propagator can be also captured by a
nonlinear relation $p=f(k)$, where $p$ and $k$ are the momentum
and the wave vector of a particle, respectively, generalizing the
commutation relation between operators $\hat{x}$ and $\hat{p}$ to
\begin{equation}\label{xp}
[\hat{x}, \hat{p}] = i\hbar \frac{\partial p}{\partial k}~\Leftrightarrow
\Delta x \Delta p \geq \frac{\hbar}{2} \left|~ \left< \frac{\partial
p}{\partial k} \right> ~\right|
\end{equation}
at quantum mechanical level \cite{Sabine}. Recently, Nouicer has
investigated the GUP effects to all orders in the Planck length on
black hole thermodynamics \cite{kn0} by arguing that the GUP up to
leading order correction in the Planck length is not enough
because the wave vector $k$ does not satisfy the asymptotic
property in the modified dispersion relation \cite{Sabine}.
Moreover, he has extended the calculation of entropy to all orders
in the Planck length for (4+1)-dimensional
Randall-Sundrum brane model \cite{kn}. Very recently, we have
extended the calculation of entropy to all orders in the Planck
length for (3+1)-dimensional Schwarzschild case by carefully
considering the entropy integral about $r$ in the range near
horizon \cite{kpn}.

In this paper, we study the entropy to all order corrections in
the Planck length of a scalar field on (2+1)-dimensional DS black
hole background carefully considering the entropy integral about
$r$ in the range $(r_H - \epsilon, r_H)$ near horizon. Contrary to
our general expectation, this study of a lower 3D case is
not so trivial in contrast to the previous result of 4D
Schwarzschild case \cite{kpn}. By using the novel equation of
states of density \cite{kn} motivated by the GUP in quantum
gravity, we calculate the quantum entropy of a massive scalar
field on (2+1)-dimensional DS black hole background introducing
the incomplete $\Gamma$-functions and carrying out numerical
calculations. As a result, we obtain the desired
Bekenstein-Hawking entropy without any artificial cutoff and
little mass approximation  satisfying the asymptotic property of
the wave vector $k$ in the modified dispersion relation. From now
on, we take the units as $G=\hbar=c=k_{B}\equiv 1$.

Let us start with the following action \cite{kpo,wlo}
\begin{equation}
\label{action}
I=\frac{1}{2\pi} \int d^3 x \sqrt{-g} \left[R -\frac{2}{l^2} \right] ,
\end{equation}
where $\Lambda= 1/{l^2}$ is a cosmological constant.
Then, the classical equation of motion yields the DS metric as
\begin{equation}
\label{metric}
ds^2 = -f(r) dt^2 +\frac{1}{f(r)} dr^2 + r^2 d \theta^2
\end{equation}
with $f(r) = (1- r^{2}/l^{2})$. The horizon is located at $r =
r_H=l$ and our spacetime is bounded by the horizon as the
two-dimensional cavity within the inner space of the horizon ($0
\le r \le l$). The inverse of Hawking temperature is given by
$\beta_H = 2\pi l$.

In this DS background, let us consider a scalar field with mass
$\mu$, which satisfies the Klein-Gordon equation
\begin{equation}
\label{kg} \frac{1}{\sqrt{-g}} \partial_{\mu}( \sqrt{-g} g^{\mu \nu}
\partial_{\nu} \Phi) - \mu^{2} \Phi = 0.
\end{equation}
Substituting the wave function $\Phi(r, \theta, t) = e^{-i\omega
t}\Psi(r, {\theta})$, we find that this equation
becomes
\begin{equation}
\label{rtheta0} \frac{\partial^{2} \Psi}{{\partial {r}}^2}  +
\left( \frac{1}{f} \frac{\partial f}{\partial r} +
\frac{1}{r}\right) \frac{\partial {\Psi}}{\partial {r}} +
\frac{1}{f} \left({\frac{1}{r^2}}\frac{{\partial}^{2}}{\partial
{\theta}^{2}} + \frac{\omega^{2}}{f} - \mu^{2} \right)\Psi = 0.
\end{equation}
By using the Wenzel-Kramers-Brillouin approximation \cite{tho}
with $\Psi \sim exp[iS(r,\theta)]$, we have
\begin{equation}
\label{wkb} {p_{r}}^{2} = \frac{1}{f}\left(\frac{\omega^{2}}{f} -
\mu^{2} - \frac{p^2_\theta}{r^2}\right),
\end{equation}
where $p_{r} = \frac{\partial S}{\partial r}$ and
 $p_{\theta} = \frac{\partial S}{\partial \theta}$.
On the other hand, we also obtain the square module momentum
\begin{equation}
\label{smom} p^{2} = p_{i}p^{i} = g^{rr}{p_{r}}^{2} + g^{\theta
\theta}{p_{\theta}}^{2} = \omega^{2}/f - \mu^{2}
\end{equation}
with the condition $\omega\geq\mu\sqrt{f}$. Moreover, considering the
modified dispersion relation given by (\ref{xp}), the usual
momentum measure $\prod^n_{i=1} dp^i$ is deformed to
\begin{equation}\label{pm}
\prod^n_{j=1} dp^j \prod^n_{i=1} \frac{\partial k^i}{\partial p^j}.
\end{equation}
For simplicity, we will restrict ourselves to
the isotropic case in one space-like dimension in the following. Then, according to
the Refs. \cite{spallucci,nouicer}, we have
\begin{equation}
\frac{\partial p}{\partial k}=~e^{\lambda p^{2}}, \label{measure}
\end{equation}%
where $\lambda$ is a dimensionless constant of order one in the
Planck length units.

Now, let us consider the deformed algebra given by $\left[
X,P\right] =i~e^{\lambda P^{2}}$ with the representations $X
\equiv i~e^{\lambda P^{2}} {\partial _{p}}$ and $P \equiv p$ of
the position and momentum operators, respectively. Then, this
algebra leads to the generalized uncertainty relation
including all order corrections in the Planck length as
\begin{equation}
\Delta X \Delta P \geq \frac{1}{2}\left\langle e^{\lambda P^{2}}
\right\rangle \geq \frac{1}{2} e^{\lambda \left( \left( \Delta
P\right)^{2}+\left\langle P\right\rangle ^{2}\right)}. \label{xp1}
\end{equation}
Note that $\left\langle P^{2n}\right\rangle \geq \left\langle
P^{2}\right\rangle ^{n}$ and $\left( \Delta
P\right)^{2}=\left\langle P^{2}\right\rangle -\left\langle
P\right\rangle^{2}$.

Next, in order to investigate the quantum implications of this
deformed algebra, let us solve the above relation (\ref{xp1}) for
$\Delta P$ that is satisfied with the equality sign. Using the
definition of the function of $W(\xi)\equiv -2\lambda \left(\Delta
P\right)^{2}$ and the Lambert function of $W\left( \xi\right)
e^{W\left( \xi\right) }=\xi$ \cite{Lambert}, we obtain the
momentum uncertainty as follows
\begin{equation}
\Delta P =\frac{e^{\lambda \langle P \rangle^2 }}{2\Delta X}
e^{\lambda \left(\Delta P\right)^{2}}. \label{argu}
\end{equation}
In order to have a real physical solution for $\Delta P$ of the
Lambert function, the argument of the Lambert function is required
to satisfy $\xi \geq -1/e$, which naturally leads to the position
uncertainty as
\begin{equation}
\Delta X \geq \sqrt{{e\lambda}/2}~e^{\lambda \langle P
\rangle^2} \equiv \Delta X _{\min}.
\end{equation}
Here, $\Delta X_{\min }$ is a minimal uncertainty in position.
Moreover, this minimal length intrinsically derived for physical
states with $\left\langle P\right\rangle =0$ is given by
\begin{equation}
\Delta X^A_{0}=\sqrt{{e\lambda}/2}, \label{min}
\end{equation}
which is the absolutely smallest uncertainty in position. In fact,
this minimal length plays a role of a brick wall cutoff
effectively giving the thickness of the thin-layer near the
horizon \cite{li,liu,kkp,kpo}. Furthermore, a series expansion of
Eq. (\ref{argu}) with $\left\langle P\right\rangle=0$ naturally
includes the well-known form of the GUP up to the leading order
correction in the Planck length units \cite{kkp,kpo} as follows
\begin{equation}
\label{gupL} \Delta X \Delta P \approx \frac{1}{2} \left[ 1+ \lambda
\left( \Delta P\right)^{2} + {\cal O}\left( \left( \Delta
P\right)^{4} \right) \right].
\end{equation}
Then, the minimal length up to the leading order is given by
$\Delta X^L_0=\sqrt{\lambda}~<~\Delta X^A_{0}$, where the
superscripts ``$L$" and ``$A$" denote the leading order and all
orders, respectively. However, only this leading order correction
of the GUP does not satisfy the property that the wave vector $k$
asymptotically reaches the cutoff in large energy region as
recently reported in Ref. \cite{Sabine}.

Now, let us calculate the statistical entropy of a scalar field on
the (2+1) DS black hole background to all orders in the Planck
length by using the GUP. When the gravity is turned on, the number
of quantum states in a volume element in phase cell space based on
the GUP in the (2+1)-dimensional DS space is given by
\begin{equation}
\label{dn} dn_A = \frac{d^2 x d^2 p}{(2\pi)^2}e^{-\lambda p^2} ,
\end{equation}
where $p^2 = p^{i}p_{i}~(i = r, \theta)$ and one quantum state
corresponding to a cell of the volume is changed from $(2\pi)^2 $
into $(2\pi)^2 e^{\lambda p^2}$ in the phase space
\cite{kn0,kn,kpn}.

From Eq. (\ref{dn}), the number of quantum states with the energy
less than $\omega$ is given by
\begin{eqnarray}
\label{Allnqs} n_{A}(\omega) &=& \frac{1}{(2\pi)^2} \int dr
d\theta
dp_{r} dp_{\theta} e^{- \lambda p^2} \nonumber   \\
&=& \frac{1}{2}\int dr \frac{r}{\sqrt{f}}
\left(\frac{{\omega}^2}{f}- \mu^{2}\right) e^{- \lambda
(\frac{{\omega}^2}{f}- \mu^{2})}.
\end{eqnarray}

On the other hand, for the bosonic case the free energy at the
inverse temperature $\beta$ is given by
\begin{equation}
\label{def} F_A = \frac{1}{\beta}\sum_K \ln \left[ 1 - e^{-\beta \omega_K} \right],
\end{equation}
where $K$ represents the set of quantum numbers. By using Eq.
(\ref{Allnqs}), the free energy can be rewritten as
\begin{eqnarray}
\label{LfreeE0}
 F_A~& \approx &~\frac{1}{\beta} \int dn_{A}(\omega) ~\ln
            \left[ 1 - e^{-\beta \omega} \right]  \nonumber   \\
   &=& - \int^{\infty}_{\mu\sqrt{f}} d \omega \frac{n_A(\omega)}{e^{\beta\omega} -1} \nonumber   \\
  &=& - \frac{1}{2} \int^{r_{H}}_{r_{H}-\epsilon} dr
\frac{r}{f\sqrt{f}} \int^{\infty}_{0} d\omega
\frac{{\omega}^2 e^{- \lambda {\frac{{\omega}^2}{f}}}}{(e^{\beta \omega} -1)}.
\end{eqnarray}
Here, we have taken the continuum limit in the first line and
integrated it by parts in the second line. Furthermore, in the
last line of Eq. (\ref{LfreeE0}), since $f \rightarrow 0$ near the
event horizon, {\it i.e.}, in the range $(r_H - \epsilon, r_H )$,
${{\omega}^2}/f - \mu^{2}$ in $n_A(\omega)$ becomes ${\omega}^2/f$
although we do not require the little mass approximation.

Moreover, we are only interested in the contribution from the just
vicinity near the horizon, $(r_H - \epsilon, r_H)$, which
corresponds to a proper distance of order of the minimal length,
$\sqrt{{e\lambda}/2}$. This is because the entropy closes to the
upper bound only in this vicinity, which it is just the vicinity
neglected by the brick wall method \cite{tho,gm,kkps,lz}. Then, we
have
\begin{eqnarray}
\label{invariant} \sqrt{\frac{e\lambda}{2}}=\int_{r_H
-\epsilon}^{r_H} \frac{dr}{\sqrt{f(r)}}
                \approx \sqrt{\frac{2\epsilon}{\kappa}},
\end{eqnarray}
where $\kappa$ is the surface gravity at the horizon of the black
hole and it is identified as
$\kappa=\frac{1}{2}\frac{df}{dr}|_{\beta =\beta_{H}} = 2\pi
{\beta_H}^{-1}$. Note that the Taylor's expansion of $f(r)$ near the
horizon is given by $f(r) \approx  2 \kappa (r_{H}-r)+ {\cal
O}\left( (r_{H}-r)^2 \right)$.

Before calculating the entropy, let us mention that Yoon et. al.
have recently suggested that since the minimal length parameter
$\lambda$ is related to the brick wall cutoff $\epsilon$ in Eq.
(\ref{invariant}), the entropy integral about $r$ in the range
near the horizon should be carefully treated for a convergent
entropy \cite{yoon}. In particular, although the term $(e^{\beta
\omega}-1)$ in Eq. (\ref{LfreeE0}) with $x=
\sqrt{\frac{\lambda}{f}}\omega$ was expanded in the previous works
giving $\beta\sqrt{\frac{f}{\lambda}}x$, one may not simply expand
up to the first order because the upper bound in the near horizon
is independent of $\epsilon$ as shown in the relation $0\leq
\frac{f}{\lambda} = \frac{2\kappa(r_H -r)}{\lambda} \leq \frac{2
\kappa \epsilon}{\lambda} = \kappa^2$~\cite{kpn}.

Now, let us carefully consider the integral about $r$ near the horizon
by extracting out the $\epsilon$-factor through the Taylor's
expansion of $f(r)$. Then, the free energy of $F_A$ in Eq.
(\ref{LfreeE0}) can be written as
\begin{equation}
\label{LfreeEnoMass}
  F_A \approx - \frac{1}{2}
   \int^{\infty}_{0} d\omega
   \frac{\omega^2}{(e^{\beta \omega} -1)} \Lambda_A(\omega,\epsilon),
\end{equation}
where $\Lambda_A(\omega,\epsilon)$ is defined by
\begin{equation}
\label{Lam}
  \Lambda_A \equiv
 \int^{r_H}_{r_{H}-\epsilon} dr \frac{r}{\left(2\kappa(r_{H}-r)\right)^{3/2}}
   e^{- \frac{\lambda {\omega}^2}{2\kappa(r_{H}-r)}}.
\end{equation}
By defining $t=\frac{\lambda {\omega}^2}{2\kappa(r_{H}-r)}$,
$\Lambda_A$ becomes
\begin{eqnarray}
\label{LfreeEnoMass1}
 \Lambda_A &=&  \frac{1}{2\kappa}
    \int^{\infty}_{\xi} dt
 \left(\frac{r_H}{\sqrt{\lambda {\omega}^2}}t^{-1/2}
    -\frac{\sqrt{\lambda {\omega}^2}}{2\kappa}t^{-3/2} \right)
   e^{- t} \nonumber\\
   &=&  \frac{r_H}{2\kappa \sqrt{\lambda\omega^2}} \Gamma\left(\frac{1}{2},~\xi\right)
   - \frac{\sqrt{\lambda\omega^2}}{(2\kappa)^2} \Gamma\left(-\frac{1}{2},~\xi\right),
\end{eqnarray}
where the incomplete $\Gamma$-functions are given as
\begin{equation}
\Gamma(a,\xi)=\int^{\infty}_{\xi}t^{a-1} e^{-t} dt, ~~\xi \equiv \frac{\lambda {\omega}^2}{2\kappa
   \epsilon}.
\end{equation}
Then, the entropy is given by
\begin{eqnarray}
\label{finalS}
  S_A &=& \beta^2 \frac{\partial F_A}{\partial\beta}\mid_{\beta=\beta_H} \nonumber \\
      & \approx & \frac{1}{4} \left(\frac{\delta_1}{2\pi^2 \sqrt{\lambda}}\right)(2\pi r_H) -
      \frac{\delta_2}{8\pi^3 r_H }\sqrt{\lambda},
\end{eqnarray}
where the numerical values $\delta_1$ and $\delta_2$ are given by
\begin{eqnarray}
\label{del}
  \delta_1 &=& \int^{\infty}_{0} dy \frac{y^2}{\sinh^{2}{y}}\Gamma\left(\frac{1}{2},~\frac{2y^2}{e \pi^2}\right)\approx 2.02,  \nonumber \\
  \delta_2 &=& \int^{\infty}_{0} dy \frac{y^4}{\sinh^{2}{y}}\Gamma\left(-\frac{1}{2},~\frac{2y^2}{e \pi^2}\right)\approx5.51.
\end{eqnarray}
This is the all order corrected finite entropy based on the GUP.
Furthermore, if we assume $\lambda = \alpha
{l_{P}}^{2}$, where $l_{P}$ is the Planck length, and in the system
of the Planck units $l_{P}= 1$, then the entropy can be rewritten by
the desired perimeter law as $S_{A} = \frac{1}{4} (2\pi r_{H})$ with
$\alpha =\frac{\delta^2_1}{4\pi^4}$ neglecting the second term,
which consists of the product of $\sqrt{\lambda}$ and $r^{-1}_H$,
in the large black hole limit.

On the other hand, in order to compare the dominant leading term in Eq. (\ref{finalS})
with that of the usual  approximation approach \cite{li,liu,kkp,kn},
let us calculate the entropy in the usual coarse-grained
approximation. In terms of the variable $x=\omega \sqrt{\lambda /f}$
\ and the fact that $e^{\beta \omega }-1=e^{\beta
\sqrt{\frac{f}{\lambda }}x}-1\approx \beta
\sqrt{\frac{f}{\lambda}}x$ for $f\rightarrow 0$, we have
\begin{eqnarray} \label{freeEf1}
 F^o_{A} & = & - \frac{1}{2\lambda\beta} \int^{r_{H}}_{r_{H}-\epsilon} dr
\frac{r}{\sqrt{f}} \int^{\infty}_{0} dx ~x ~e^{-x^2} \nonumber \\
& = & - \frac{1}{4 \lambda\beta} \int^{r_{H}}_{r_{H}-\epsilon} dr
\frac{r}{\sqrt{f}}.
\end{eqnarray}
Then, when $r\rightarrow r_{H}$, we get the entropy to all orders from the free
energy (\ref{freeEf1}) as follows
\begin{eqnarray} \label{Aentropy0}
S^o_{A} &=& \beta^2 \frac{\partial F^o_A}{\partial \beta}|_{\beta =\beta_{H}} \nonumber \\
&=& \frac{1}{4\lambda} \int^{r_{H}}_{r_{H}-\epsilon} dr \frac{1}{\sqrt{f}}~r
 \approx  \frac{1}{4\lambda} \sqrt{\frac{e\lambda}{2}} ~r_H  \nonumber \\
&=& \frac{1}{4} \left(\frac{\sqrt{e}}{2\sqrt{2}
\pi}\frac{1}{\sqrt{\lambda}}\right)(2\pi r_H).
\end{eqnarray}
Note that if we assume $\lambda = \alpha {l_{P}}^{2}$, where
$l_{P}$ is the Planck length, and in the system of the Planck units
$l_{P}= 1$, then the entropy can be also rewritten by the desired
perimeter law as $S^o_{A} = \frac{1}{4} (2\pi r_{H})$ with $\alpha
=\frac{e}{8\pi^2}$.

Finally, it seems to be appropriate to comment on the dominant leading
term of the entropy (\ref{finalS}), which is obtained through the
Taylor expansion of $f(r)$ near the horizon, comparing with the
entropy (\ref{Aentropy0}), which is obtained through the usual
approximation approach. Although their values are different, we
have obtained the desired Bekenstein-Hawking entropy by properly
adjusting the minimal length parameter $\lambda$ for both
approximation approaches. Therefore, we may propose that although
by carefully considering all order corrected GUP in the Planck
length the entropy can be finely obtained, it is enough to
practically use the simple usual approximation for the desired
entropy due to the existence of the adjustable parameter $\alpha$.

In summary, by using the generalized uncertainty principle, we
have investigated the entropy to all orders in the Planck length
of the massive scalar field on the (2+1)-dimensional de Sitter
black hole background carefully considering the entropy integral
about $r$ in the range $(r_H - \epsilon, r_H )$ near the horizon.
As a result, we have finely obtained the desired Bekenstein-Hawking
entropy without any artificial cutoff and any little mass
approximation satisfying the asymptotic property of the wave
vector $k$ in the modified dispersion relation.

\begin{acknowledgments}
Y.-W. Kim was supported by the
Korea Research Foundation Grant funded by Korea Government
(MOEHRD): KRF-2007-359-C00007. Y.-J. Park was supported by
the Korea Science and Engineering Foundation (KOSEF) grant
funded by the Korea government (MOST) (R01-2007-000-20062-0).

\end{acknowledgments}


\begin{references}

\bibitem{bek} J. D. Bekenstein, Phys. Rev. D {\bf 7}, 2333 (1973); Phys. Rev. D {\bf 9}, 3292
(1974).

\bibitem{haw} S. W. Hawking, Commun. Math. Phys. {\bf 43}, 199 (1975).

\bibitem{tho} G. 't Hooft, Nucl. Phys. B {\bf 256}, 727 (1985).

\bibitem{gm} A. Ghosh and P. Mitra, Phys. Rev. Lett. {\bf 73}, 2521 (1994); S. P.
de Alwis and N. Ohta, Phys. Rev. D {\bf 52}, 3529 (1995); R.-G. Cai
and Y.-Z. Zhang, Mod. Phys. Lett. A {\bf 11}, 2027 (1996); S.P. Kim,
S.K. Kim, K.-S. Soh, and J.H. Yee, Phys. Rev. D {\bf 55}, 2159
(1997); J. Jing and M.-L. Yan, Phys. Rev. D {\bf
60}, 084015 (1999); M. Kenmoku, K. Ishimoto, K.K. Nandi, and K.
Shigemoto, Phys. Rev. D {\bf 73}, 064004 (2006).

\bibitem{kkps} J.-W. Ho, W. Kim, Y.-J. Park,
and H.-J. Shin, Class. Quantum Grav. {\bf 14}, 2617 (1997);
W. Kim, J.J. Oh, and Y. J. Park, Phys. Lett. B {\bf 512}, 131 (2001).

\bibitem{lz} X. Li and Z. Zhao, Phys. Rev. D {\bf 62}, 104001 (2000);
F. He, Z. Zhao, and S-W. Kim, Phys. Rev. D {\bf 64}, 044025
(2001); C.-J. Gao and Y.-G. Shen, Phys. Rev. D {\bf 65}, 084043
(2002).

\bibitem{li} X. Li, Phys. Lett. B {\bf 540}, 9 (2002);
X. Sun and W. Liu, Mod. Phys. Lett. A {\bf 19}, 677 (2004).

\bibitem{liu}  R. Zhao, Y. Q. Wu, and L. C. Zhang, Class. Quantum
Grav. {\bf 20}, 4885 (2003); W. B. Liu, Chin. Phys. Lett. {\bf 20}, 440 (2003);
C. Liu, X. Li, and Z. Zhao, Gen. Rel. Grav. {\bf 36}, 1135 (2004);
C.-Z. Liu, Int. J. Theor. Phys. {\bf 44}, 567 (2005).

\bibitem{kkp} W. Kim, Y.-W. Kim, and Y.-J. Park,
Phys. Rev. D {\bf 74}, 104001 (2006); W. Kim, Y.-W. Kim, and Y.-J.
Park, Phys. Rev. D {\bf 75}, 127501 (2007).

\bibitem{gup} A. Kempf, G. Mangano, and R. B. Mann, Phys. Rev. D {\bf 52}, 1108 (1995);
L. J. Garay, Int. J. Mod. Phys. A {\bf 10}, 145 (1995); L. N. Chang,
D. Minic, N. Okamura, and T. Takeuchi, Phys. Rev. D {\bf 65}, 125028
(2002).

\bibitem{gup1} F. Scardigli and R. Casadio, Class. Quant. Grav. {\bf 20}, 3915
(2003); A.J.M. Medved and E. C. Vagenas, Phys. Rev. D {\bf 70}, 124021 (2004);
Y. Ling, B. Hu, and X. Li, Phys. Rev. D {\bf 73}, 087702 (2006);
Y. Ko, S. Lee, and S. Nam, [{\tt arXiv:hep-th/0608016}];
Y.S. Myung, Y.-W. Kim, and Y.-J. Park, Phys. Lett. B {\bf 645}, 393 (2007).

\bibitem{kpo} W. Kim, Y.-W. Kim, and Y.-J. Park, J. Korean Phys. Soc. {\bf
49}, 1360 (2006).

\bibitem{wlo} W. T. Kim, Phys. Rev. {\bf D59}, 047503 (1999);
A. Lopez-Ortega, Gen. Rel. Grav. {\bf 36}, 1299 (2004).

\bibitem{yoon} M. Yoon, J. Ha, and W. Kim,  Phys. Rev. D {\bf 76}, 047501 (2007).

\bibitem{kn0} Kh. Nouicer, Phys. Lett. B {\bf 646}, 63 (2007).

\bibitem{kn} Kh. Nouicer, [{\tt arXiv:gr-qc/0705.2733}].

\bibitem{kpn} Y.-W. Kim and Y.-J. Park, Phys. Lett. B (in press) [{\tt arXiv:gr-qc/0707.2128v2}].

\bibitem{Moffat} J. Moffat, Phys. Lett. B {\bf 506}, 193 (2001).

\bibitem{spallucci} A. Smailagic and E. Spallucci, J. Phys. A {\bf 36}, L467
(2003); A. Smailagic an d E. Spallucci, J. Phys. A {\bf 37}, 7169
(2004).

\bibitem{nouicer} Kh. Nouicer and M. Debbabi, Phys. Lett. A {\bf 361}, 305 (2007).

\bibitem{ven} G. Veneziano, Europhys. Lett. {\bf 2}, 199 (1986);
    D. J. Gross and P. F Mende, Phys. Lett. B {\bf 197}, 129 (1987).

\bibitem{mag} T. Padmanabhan, Class. Quant. Grav. {\bf 4}, L107 (1987); M. Maggiore, Phys. Lett. B
     {\bf 304}, 65 (1993); F. Scardigli, Phys. Lett. B {\bf 452}, 39, (1999).

\bibitem{Cam} G. Amelino-Camelia, M. Arzano, Y. Ling, and G. Mandanici,
  Class. Quantum Grav. {\bf 23}, 2585 (2006); K.
Nozari, Phys. Lett. B {\bf 635}, 156 (2006).

\bibitem{Sabine} S. Hossenfelder, Phys. Rev. D {\bf 73}, 105013 (2006);
Class. Quant. Grav. {\bf 23}, 1815 (2006).

\bibitem{fsp} M. Fontanini, E. Spallucci, and T. Padmanabhan, Phys. Lett. B {\bf 633},
   627 (2006).

\bibitem{Lambert} J. Matyjasek, Phys.
Rev. D {\bf 70} 047504 (2004); Y. S. Myung, Y.-W. Kim, and Y.-J.
Park, [{\tt arXiv:gr-qc/0705.2478}].

\end{references}
\end{document}